%% file: main.tex
\documentclass[prb, aps, twocolumn, superscriptaddress, longbibliography]{revtex4-2}

\usepackage[utf8]{inputenc}
\usepackage{amsmath,amsfonts,amssymb}
\usepackage{fancyhdr}
\usepackage{textcomp}
\usepackage{graphicx}
\usepackage[colorlinks=true,linkcolor=blue,citecolor=blue,urlcolor=blue]{hyperref}
\usepackage{ulem}

\fancypagestyle{firstpage}{%
  \fancyhf{}%
  \fancyfoot[R]{%
    \scriptsize
    \scalebox{0.8}{\textcircled{\scriptsize c}}\,2025\ American Physical Society%
  }%
}

\begin{document}

\title{Anomalies in G and 2D Raman modes of twisted bilayer graphene near the magic angle}

\author{Darshit Solanki}
\affiliation{Indian Institute of Science, Bangalore 560012, India}

\author{Kenji Watanabe}
\affiliation{National Institute for Materials Science, Tsukuba 305-0044, Japan}

\author{Takashi Taniguchi}
\affiliation{National Institute for Materials Science, Tsukuba 305-0044, Japan}

\author{A. K. Sood}
\email{asood@iisc.ac.in}
\affiliation{Indian Institute of Science, Bangalore 560012, India}

\author{Anindya Das}
\email{anindya@iisc.ac.in}
\affiliation{Indian Institute of Science, Bangalore 560012, India}

\begin{abstract}
    The role of twist angle ($\theta_t$) in tailoring the physical properties of heterostructures is emerging as a new paradigm in two-dimensional materials. The influence of flat electronic bands near the magic angle ($\sim$1.1$^{\circ}$) on the phononic properties of twisted bilayer graphene (t\mbox{-}BLG) is not well understood. In this work, we systematically investigate the G and 2D Raman modes of t\mbox{-}BLG samples with twist angles ranging from $\sim$0.3$^{\circ}$ to $\sim$3$^{\circ}$ using micro-Raman spectroscopy. A key finding of our work is the splitting of the G mode near the magic angle due to moir\'{e} potential induced phonon hybridization. The linewidth of the low-frequency component of the G mode (G$^-$), as well as the main component of the 2D mode, exhibits enhanced broadening near the magic angle due to increased electron-phonon coupling, driven by the emergence of flat electronic bands. Additionally, temperature-dependent Raman measurements (6–300 K) of magic-angle twisted bilayer graphene sample ($\theta_t \sim$ 1$^{\circ}$) reveal an almost tenfold increase in phonon anharmonicity-induced temperature variation in both components of the split G mode, as compared to Bernal-stacked bilayer graphene sample, further emphasizing the role of phonon hybridization in this system. These studies could be important for understanding the thermal properties of the twisted bilayer graphene systems.
\end{abstract}

\maketitle
\thispagestyle{firstpage}

\section{Introduction}

    A superlattice forms when two identical periodic structures are superposed with a slightly mismatched rotation angle, a phenomenon commonly observed in two-dimensional materials. This twist in the top layer relative to the bottom layer creates a moir\'{e} superlattice, introducing an additional periodic potential known as the moir\'{e} potential \cite{ohta2012evidence}. The moir\'{e} potential originating from atomic reconstruction could significantly influence the properties of twisted bilayer graphene (t\mbox{-}BLG) near the magic angle. Therefore, such controlled modifications in twisted bilayer graphene systems offer opportunities to tailor their physical properties, leading to novel phenomena \cite{superconductivity1, mesple2021heterostrain, superconductivity2, oh2021evidence, lu2019superconductors, novoselov2006unconventional, grimvall1976electron}. Notably, at small twist angles ($\theta_t$), particularly around the magic angle ($\theta_t \approx$ 1.1$^{\circ}$), t\mbox{-}BLG exhibits a flat band structure \cite{superconductivity1, mesple2021heterostrain}, resulting in exceptional properties such as superconductivity \cite{superconductivity1, superconductivity2, oh2021evidence, lu2019superconductors}, anomalous quantum Hall effects \cite{novoselov2006unconventional}, transport mediated by relativistic massless Dirac fermions, and high electrical and thermal conductivities \cite{superconductivity1, superconductivity2}. Additionally, atomic reconstruction-induced domain structures have been observed in t\mbox{-}BLG at small twist angles \cite{absentMPabove2deg-1, woods2014commensurate, kim2017evidence, nam2017lattice, gargiulo2017structural, zhang2018structural, novoselov20162d}, making small-angle t\mbox{-}BLG a compelling system to study \cite{alden2013strain, lin2018shear}.

    Despite extensive investigations into graphene and its twisted bilayer counterparts, a comprehensive understanding of its electronic, phononic, and optical properties near the magic angle remains incomplete. These properties are crucial for designing devices customized for various applications. Therefore, in this work, we aimed to investigate the behavior of the G and 2D Raman modes of t\mbox{-}BLG, as these modes offer valuable insights into its electronic and phononic properties. To achieve this, we systematically fabricated multiple t\mbox{-}BLG samples with twist angles ($\theta_t$) ranging from approximately 0.3$^{\circ}$ to 3$^{\circ}$. We employed micro-Raman spectroscopy to investigate different Raman active modes of t\mbox{-}BLG over this range of twist angles. Raman spectroscopy, a pivotal tool in the structural characterization of graphitic materials \cite{dresselhaus2010perspectives, ferrari2013raman}, is particularly useful for understanding the complex interactions between electrons and phonons in graphene. In t\mbox{-}BLG, the energy separation between nearly flat bands and the nearest conduction and valence bands aligns closely with the photon energy used in optical investigations. This alignment makes t\mbox{-}BLG an interesting subject for optical studies, with its properties intricately linked to variations in the electronic band structure and van Hove singularities (VHS), which are strongly dependent on the twist angle.

    We further explored the Raman spectra of a t\mbox{-}BLG sample near magic angle as a function of temperature (ranging from 6 K to 300 K) to understand phonon anharmonicities. Our analysis reveals the evolution of the G mode (also known as E$_{2g}$ mode) and the 2D mode with respect to the twist angle, driven primarily by the lattice reconstruction-induced moir\'{e} potential \cite{darshit2024twist}. Noteworthy trends include a maximum in the full width at half maximum (FWHM) of the G mode at $\theta_t \sim$ 1$^{\circ}$. Additionally, we observed the splitting of the G mode near the magic angle. Our findings provide crucial insights into the behavior of phonons in t\mbox{-}BLG near the magic angle, offering guidance for the design and fabrication of novel devices.

\section{Experimental Methods}

    \begin{figure*}
    \centering
    \includegraphics[width=0.7\textwidth]{{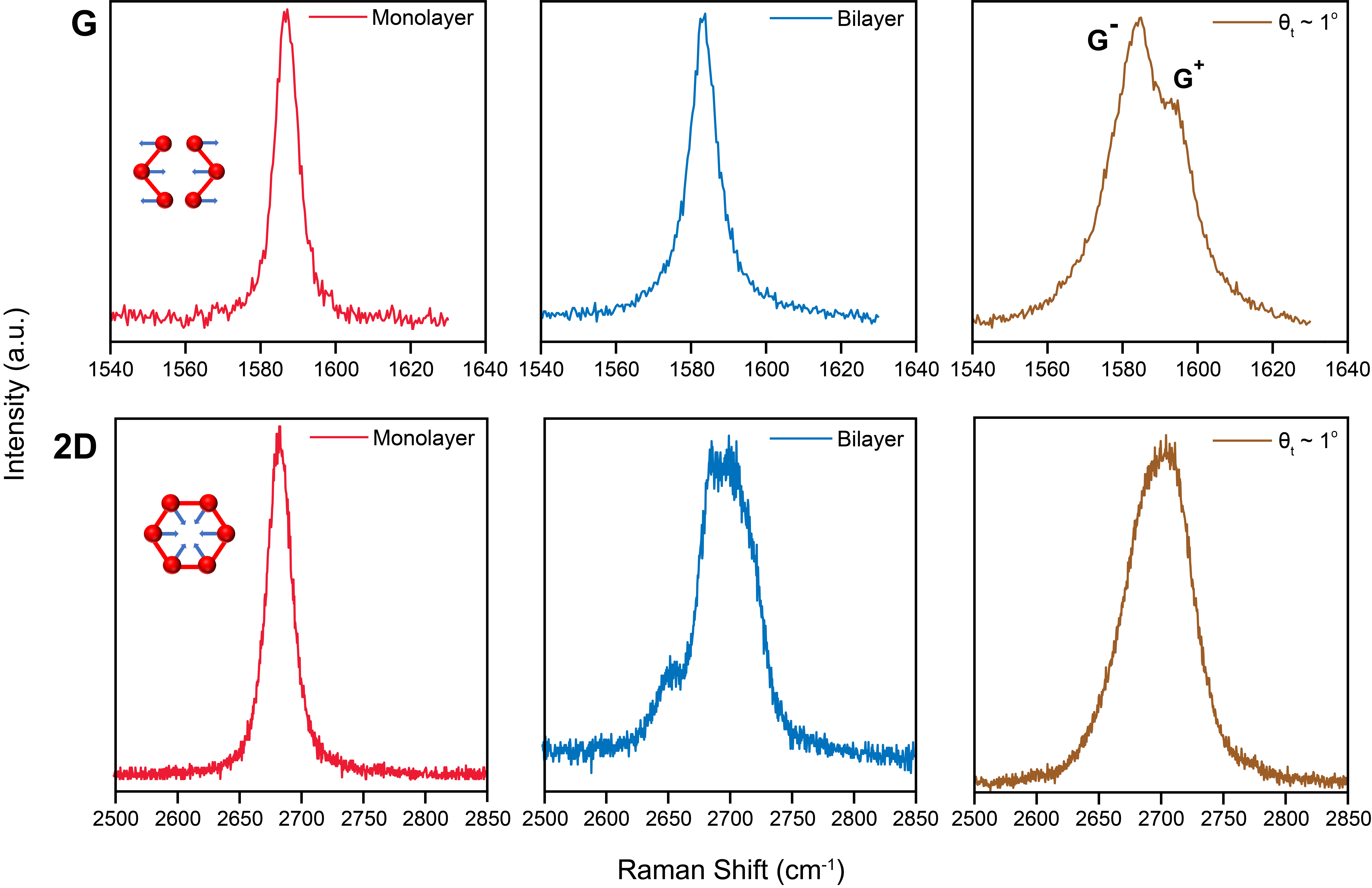}}
    \caption{G and 2D Raman modes for three different samples: monolayer graphene, bilayer graphene, and t\mbox{-}BLG with a twist angle of $\sim$1$^{\circ}$. The upper panel illustrates the G mode, while the lower panel displays the 2D mode for each respective sample. Insets show schematic representations of the vibrational modes corresponding to the G and 2D Raman bands.}
    \label{Fig1}
    \end{figure*}

    Graphene flakes were exfoliated onto a silicon substrate using the conventional mechanical exfoliation method \cite{huang2020universal, gao2018mechanical, yi2015review}, which facilitates the isolation of single graphene layers. Post-exfoliation, an optical microscope was employed to examine the graphene flakes and verify the presence of monolayer graphene by evaluating their visual contrast. Once identified, Raman spectroscopy was performed to confirm the number of graphene layers.

    Subsequently, twisted bilayer graphene samples were fabricated using the standard tear-and-stack technique \cite{kim2016van, guo2021stacking} (details provided in the Supplementary Information). To preserve the pristine quality of the fabricated t\mbox{-}BLG samples, they were encapsulated within hexagonal boron nitride (hBN) flakes. This encapsulation acts as a protective barrier, preventing potential contamination and degradation \cite{holler2019air}. The hBN flakes, with their atomically flat surfaces and inert chemical properties \cite{zhang2017two, taychatanapat2011quantum}, provide an ideal environment for maintaining and improving the electronic and structural characteristics of the t\mbox{-}BLG samples \cite{dean2010boron, xue2011scanning, decker2011local}.

    \begin{figure*}
    \centering
    \includegraphics[width=\textwidth]{{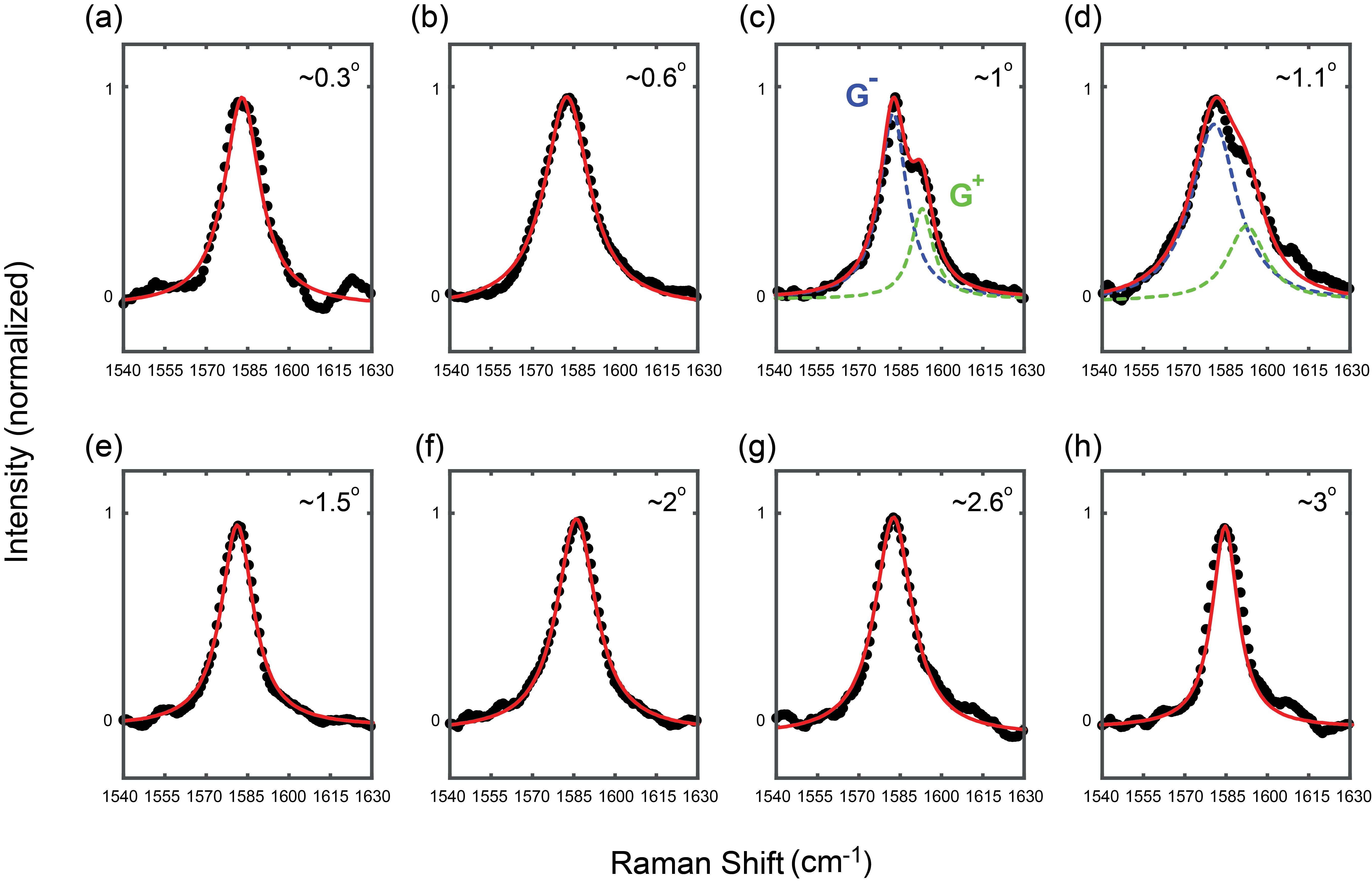}}
    \caption{The figure shows the G mode of t\mbox{-}BLG at various angles ranging from $\sim$0.3$^{\circ}$ to $\sim$3$^{\circ}$. Experimental data is represented by black dots, while the red lines show the Lorentzian fits to the data. The blue and green dotted lines in (c) and (d) show the individual Lorentzian fits.}
    \label{G-evol}
    \end{figure*}

    \begin{figure*}
    \centering
    \includegraphics[width=\textwidth]{{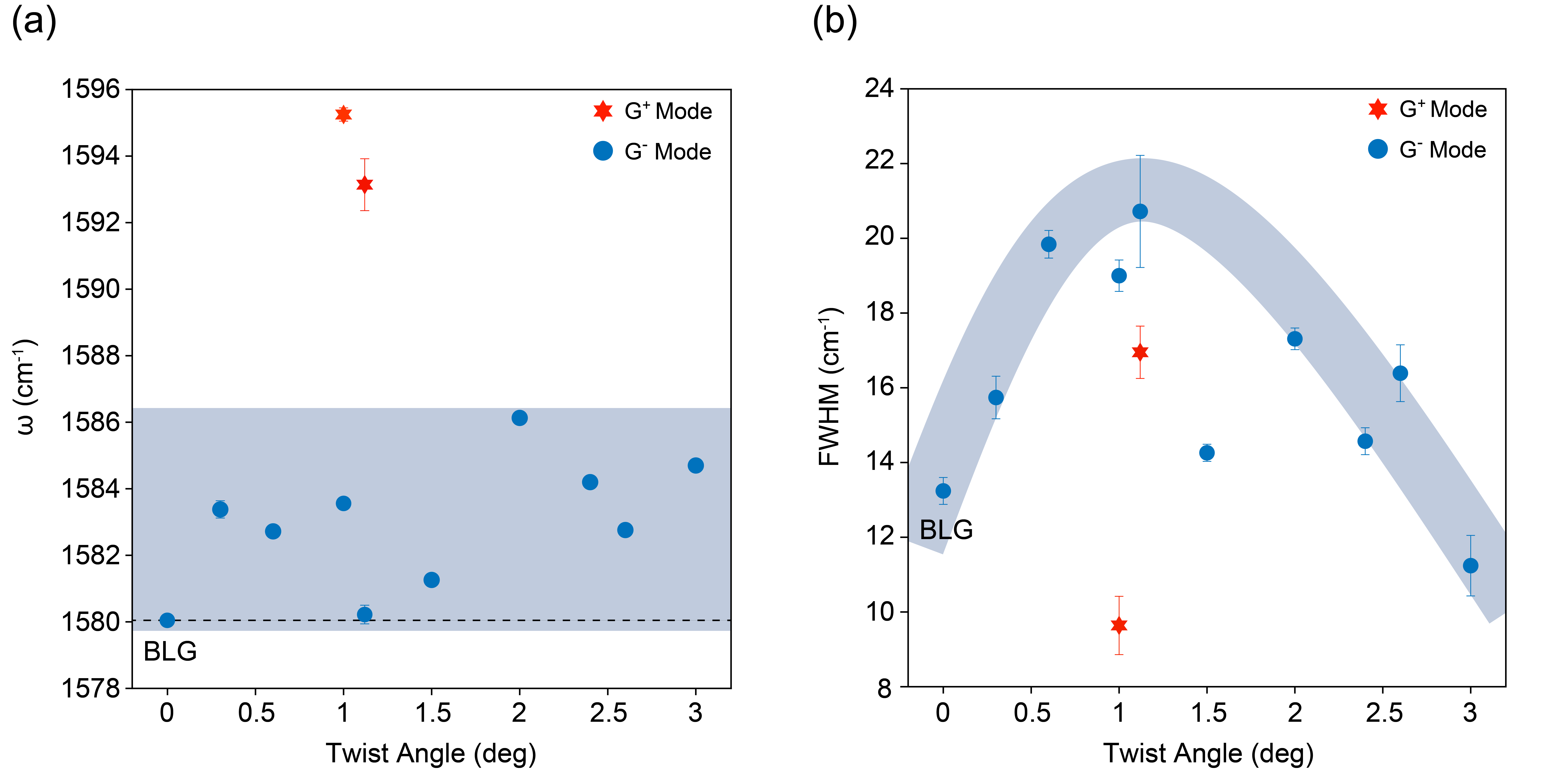}}
    \caption{Frequencies and linewidths of the G$^+$ and G$^-$ modes as functions of twist angle. The grey bands in the figures serve as visual guides.}
    \label{G-FWHM-evol}
    \end{figure*}

    \begin{figure*}
    \centering
    \includegraphics[width=\textwidth]{{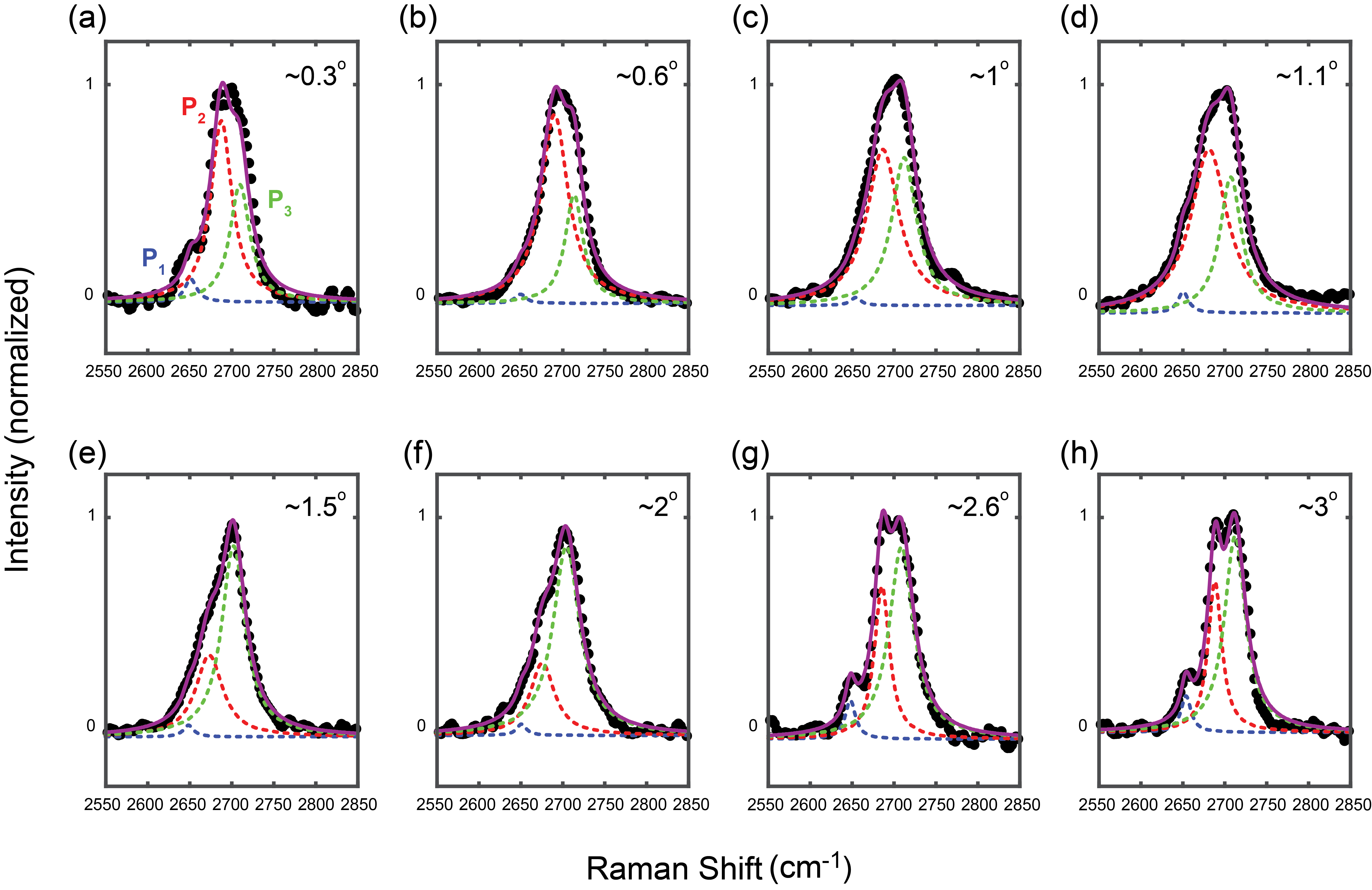}}
    \caption{The 2D mode of twisted bilayer graphene is shown for various twist angles ranging from $\sim$0.3$^{\circ}$ to $\sim$3$^{\circ}$. Experimental data are represented by black dots. The dotted lines indicate the individual Lorentzian fits, and the solid violet line shows the overall fit.}
    \label{2D-evol}
    \end{figure*}

    \begin{figure*}
    \centering
    \includegraphics[width=\textwidth]{{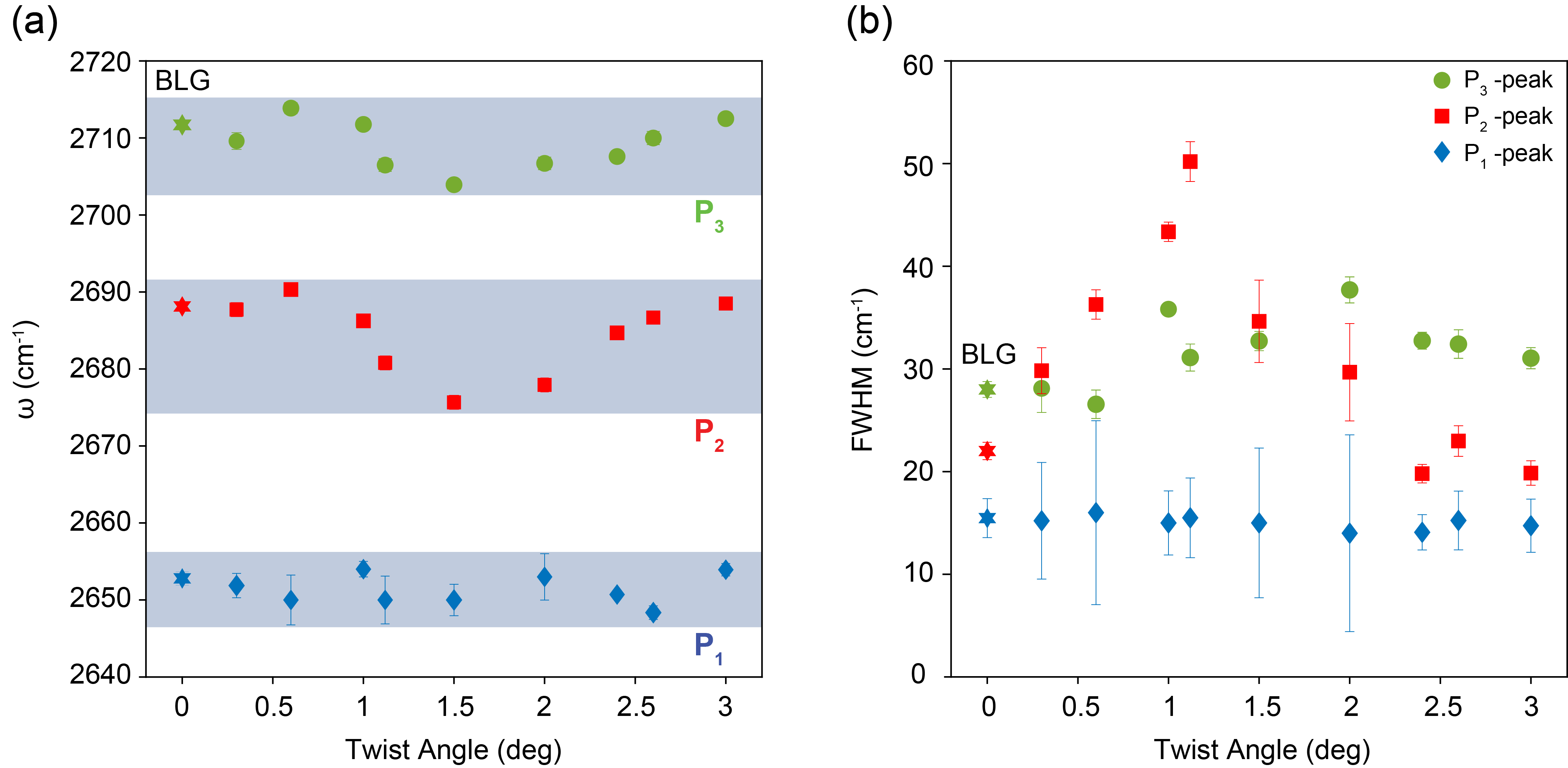}}
    \caption{Frequency (Panel a) and FWHM (Panel b) of the 2D sub-peaks in twisted bilayer graphene are plotted as functions of twist angle. Data points corresponding to 0$^{\circ}$ represent the values for Bernal-stacked bilayer graphene. The grey band in the figure serves as a visual guide.}
    \label{2D-FWHM-evol}
    \end{figure*}

    \begin{figure*}
    \centering
    \includegraphics[width=0.7\textwidth]{{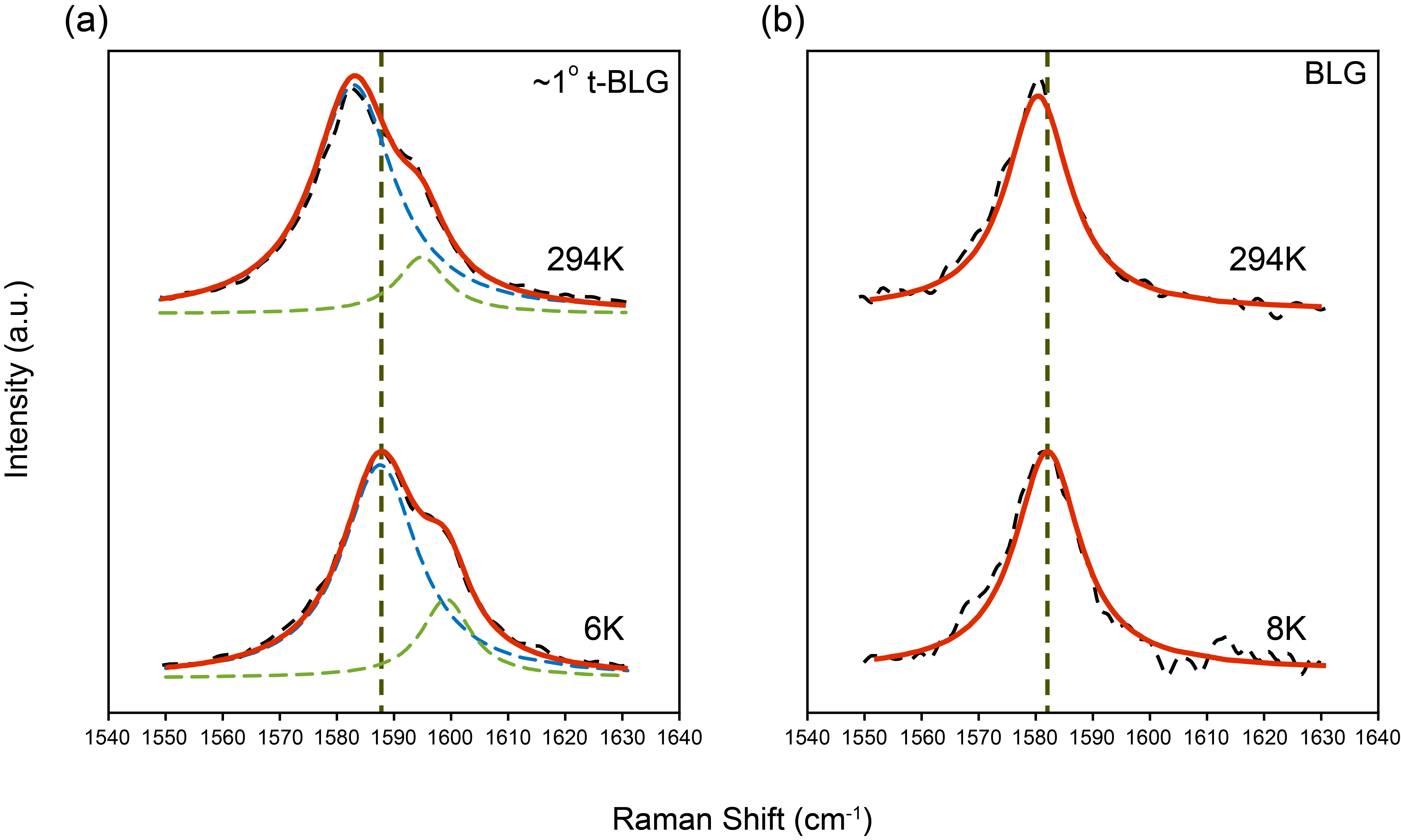}}
    \caption{Raman spectra of the G$^+$ and G$^-$ modes in $\sim$1$^{\circ}$ twisted bilayer graphene (t-BLG) (a), and the G mode in Bernal-stacked bilayer graphene (BLG) (b), measured at both low temperature (6K / 8K) and room temperature (294K). The black dotted lines represent the experimental data. The blue and green dashed lines show the individual Lorentzian components, and the solid red line corresponds to their sum representing the overall fit.}
    \label{Fig6}
    \end{figure*}

    \begin{figure*}
    \centering
    \includegraphics[width=0.5\textwidth]{{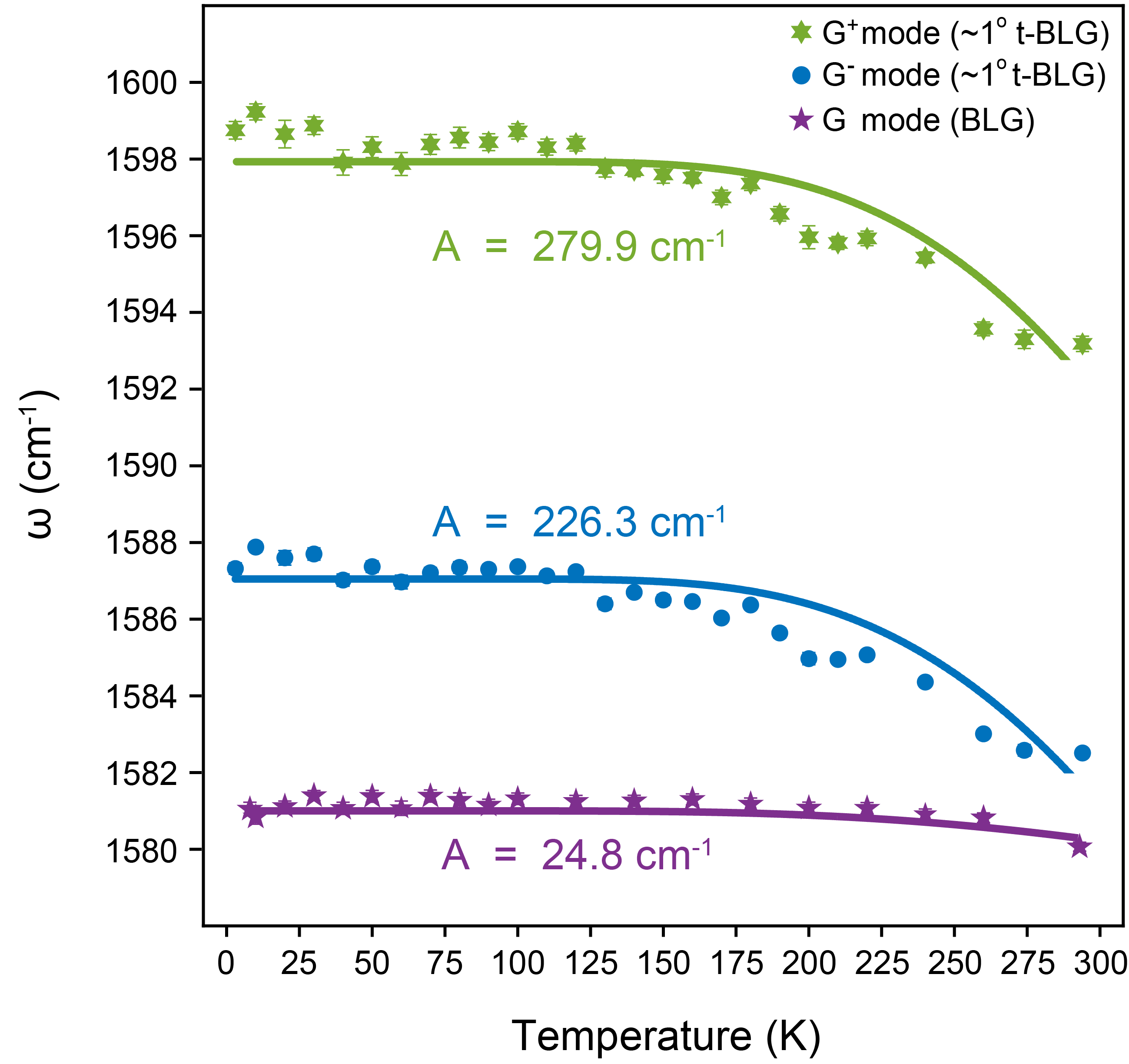}}
    \caption{The temperature dependence of the G$^+$ and G$^-$ mode frequencies in $\sim$1$^{\circ}$ t-BLG and the G mode frequency in Bernal-stacked BLG is shown. The data points represent experimental measurements, while the solid lines correspond to fits using the anharmonic model (Eq. \ref{anharmonic-eq}). The extracted anharmonic coefficients are given next to the respective datasets.}
    \label{Fig7}
    \end{figure*}

    Raman spectra of the fabricated t\mbox{-}BLG structures were recorded using a Horiba LabRAM HR spectrometer, combined with a Montana cryostat capable of maintaining an ultra-low base pressure ranging from 10$^{-6}$ to 10$^{-8}$ Torr. Raman spectra were acquired at both room temperature and low temperatures using a 532 nm excitation laser and a grating with 1800 lines/mm was used. The system has a resolution of around 0.55 cm$^{-1}$ and the laser power was kept below 1 mW to avoid laser-induced heating. For low-temperature spectra acquisition, a 50$\times$ lens with a numerical aperture (NA) of 0.50 was used, while room-temperature spectra were captured using a 100$\times$ lens with an NA of 0.90.

\section{Observations and Discussion}

    Moir\'{e} structures in twisted bilayer graphene (t\mbox{-}BLG) exhibit diverse atomic stacking configurations depending on the relative twist angles, which alters the interlayer van der Waals forces. These forces play a significant role in determining the atomic arrangements, contributing to the formation of distinct high-symmetry stacking configurations such as AA, AB, and SP (saddle point), where SP is the transition region between the AA and AB, representing the saddle points of the van der Waals energy landscape \cite{gadelha2021localization}. These stacking regions vary spatially across the graphene superlattice. When the system undergoes atomic relaxation, where atoms move to minimize the total energy, a reconstruction occurs at the atomic level. During this process, the locally stacked regions evolve towards their true minimum energy configurations, leading to more stable and energetically favorable structures \cite{alden2013strain, lin2018shear}. Additionally, atomic reconstruction-induced domain structures have been observed in t\mbox{-}BLG at small twist angles, whereas angles exceeding 2$^{\circ}$ do not show substantial reconstruction \cite{absentMPabove2deg-1,absentMPabove2deg-2}. These distinct stacking regions significantly modulate the electronic and phononic properties of the graphene. Such modulation can result in novel properties not present in the individual graphene layers or Bernal (AB) bilayers. Consequently, by precisely controlling the twist angle, it is possible to fine-tune the properties of t\mbox{-}BLG. It is important to note that the label E$_{2g}$ is strictly valid only for monolayer graphene, where the system possesses higher symmetry. In twisted bilayer graphene, the introduction of a twist angle lowers the crystal symmetry thus affecting the symmetry classification of phonon modes \cite{gontijo2024resonant}. The vibrational modes in twisted bilayer graphene cannot be straightforwardly classified with the symmetry labels used for monolayer graphene. Therefore, we simply refer to the Raman-active mode in t-BLG around 1583 cm$^{-1}$ as the G mode.
     
    The quality and defect-free nature of the graphene layers used in our samples was confirmed using Raman spectroscopy. Figure S1 of Supporting Information (SI) \cite{supplement} shows a representative Raman spectrum of a monolayer graphene sample, showcasing the absence of the characteristic D mode (disorder-induced mode) at approximately 1350 cm$^{-1}$, indicative of pristine, defect-free monolayer graphene.

    In Figure \ref{Fig1}, we show the G and 2D modes of monolayer, bilayer and $\sim$1$^{\circ}$ t\mbox{-}BLG for comparison. As can be seen that, the spectrum is distinct for monolayer, bilayer and $\sim$1$^{\circ}$ t\mbox{-}BLG samples. Monolayer and bilayer have single G peak in contrast to split G peak for $\sim$1$^{\circ}$ t\mbox{-}BLG. Similarly, a sharp single 2D peak is seen for single layer graphene (SLG) and several peaks (four) for BLG, but for $\sim$1$^{\circ}$ t\mbox{-}BLG broadened 2D peak is seen. In Figure \ref{G-evol}, we demonstrate the evolution of G mode of t\mbox{-}BLG for twist angles spanning from $\sim$0.3$^{\circ}$ to $\sim$3$^{\circ}$. The G mode, occurring at approximately 1583 cm$^{-1}$, corresponds to the E$_{2g}$ mode at the $\Gamma$-point for monolayer graphene. This mode is present in all sp$^2$ carbon systems and results from the stretching of the C-C bond in graphitic materials. Due to its high sensitivity to strain effects in sp$^2$ systems, the E$_{2g}$ mode is valuable for investigating modifications on the flat surface of graphene. The spectra have been normalized for direct comparison and analysis. The G mode of all samples, except those with twist angles of $\sim$1$^{\circ}$ and $\sim$1.1$^{\circ}$, are fitted with a single Lorentzian function to extract the peak position and full width at half maximum (FWHM). Conversely, the samples with twist angles of approximately 1$^{\circ}$ and 1.1$^{\circ}$ are fitted with two Lorentzian functions to account for peak splitting. The blue and green dotted lines represent the individual Lorentzian fittings for the split peaks, while the red solid line represents the sum of these two. For the remaining plots, the red solid line represents the single Lorentzian fitting.

    The splitting of the G mode into a doublet around the magic angle could be driven by the hybridization of phonon modes due to the moir\'{e} potential, similar to a recent study on t-WSe$_2$ \cite{darshit2024twist}. Accordingly, we infer that the G mode can split due to underlying moir\'{e} potential, resulting from the reduced symmetry of the moir\'{e} unit cell. This splitting of the G mode results in a high-frequency (G$^+$) component and a low-frequency (G$^-$) component. The moir\'{e} potential strengthens as the twist angle approaches the magic angle, leading to an increased splitting of the G mode as seen in a recent theoretical study on t\mbox{-}BLG \cite{mandal2024phonon}. It is worth noting that the observed splitting of the G mode may be partially influenced by strain arising from atomic reconstruction, similar to what has been reported for twisted WSe$_2$ \cite{darshit2024twist}. However, separating the contributions from phonon hybridization and strain remains challenging and requires further theoretical investigations.
    
    We further rule out the possibility that the observed splitting originates from either an intralayer resonance process \cite{eliel2018intralayer, moutinho2021resonance} or asymmetric doping between the top and bottom graphene layers \cite{mafra2012characterizing, chung2015optical}. Regarding the intralayer resonance, the laser energy used in our measurements (2.33 eV, 532 nm) far exceeds the predicted resonance energy ($\sim$0.40 eV) required to activate the L$_a$ mode from the LO phonon branch in magic-angle twisted bilayer graphene ($\theta_t$ $\sim$ 1$^{\circ}$) \cite{eliel2018intralayer, moutinho2021resonance}, thereby ruling out the intralayer resonance as the origin. In the case of asymmetric doping, our samples are fully encapsulated in hBN, which helps maintain their pristine condition, thereby minimizing any doping imbalance. Furthermore, as shown in the Supporting Information, the Raman spectra measured from the top and bottom monolayer graphene regions adjacent to the twisted area are nearly identical, further ruling out any significant asymmetric doping.

    Figure \ref{G-FWHM-evol} shows the frequency and full width at half maximum (FWHM) of the G mode as a function of the twist angle for all the samples. No discernable trend or correlation is observed between the twist angle and the frequency of the G mode, indicating that the twist angle has no significant influence on this characteristic. However, the FWHM of the G$^-$ mode shows a clear dependence on the twist angle: it increases with increasing twist angle, peaks near 1$^{\circ}$, and then decreases as the twist angle increases further.
    
    A similar dependence of the FWHM of the G mode \cite{gadelha2021localization, barbosa2022raman} was observed in a study by Barbosa \textit{et al.} \cite{barbosa2022raman} on t\mbox{-}BLG, where they attributed the broadening to enhanced electron-phonon coupling at the magic angle caused by an increased electronic density of states (DOS). However, they did not observe any splitting of the G mode. In contrast, our data show that instead of a simple broadening, the G mode splits into two distinct components: G$^-$ (lower frequency) and G$^+$ (higher frequency). This splitting is evident in Figure \ref{G-FWHM-evol}b, where the G$^-$ mode exhibits a larger FWHM compared to the G$^+$ mode.
    
    The 2D mode observed around $\sim$2700 cm$^{-1}$ (Figure S1 in SI) arises from a second-order two-phonon process. Figure \ref{2D-evol} illustrates how the shape of the 2D mode evolves with systematic variations in twist angle. At the smallest twist angle of $\sim$0.3$^{\circ}$, the 2D mode resembles the profile of Bernal-stacked bilayer graphene. However, as the twist angle approaches the magic angle, the characteristic shoulder peak (marked P$_1$ in Figure \ref{2D-evol}a) of Bernal-stacked bilayer graphene appears to diminish, resulting in a broadened single-peak shape of the 2D mode. The apparent disappearance of the P$_1$ peak could be caused by an increase in the linewidths of the P$_1$, P$_2$, and P$_3$ peaks. This linewidth increase may result from stronger electron-phonon coupling associated with the flat electronic bands near the magic angle. Due to this broadening, the three peaks merge into a single broadened peak shape for the 2D mode, creating the impression that the P$_1$ peak has disappeared and also blurring the separation between the individual components. Although Figure \ref{2D-FWHM-evol}a shows no systematic change in the frequencies of the P$_1$, P$_2$, and P$_3$ peaks as a function of twist angle, Figure \ref{2D-FWHM-evol}b reveals a significant increase in the full width at half maximum (FWHM), particularly for the P$_2$ peak. This may be due to the complexity of fitting the 2D peak, where the majority of the broadening is likely absorbed by the P$_2$ component during fitting. This could explain the relatively unchanged FWHMs of the P$_1$ and P$_3$ peaks.
    
    Interestingly, at twist angles exceeding 2$^{\circ}$, the shoulder peak P$_1$ reappears, and the 2D band shape is similar to that of Bernal-stacked bilayer graphene. This resurgence may occur because the flat bands are not present beyond the magic angle, reducing the linewidths. This finding contrasts Ref. \cite{barbosa2022raman} showing a single Lorentzian-shaped 2D band around 3$^{\circ}$ twist angles, attributed to the predominance of SP/AA regions in the resulting moir\'{e} pattern, with negligible contribution from AB/BA regions. It may be noted, however, that high-symmetry regions such as AB/BA, SP, and AA are generally assumed to be absent beyond 2$^{\circ}$ in t\mbox{-}BLG \cite{absentMPabove2deg-1, absentMPabove2deg-2}.

    In order to explore how twist angle-dependent phonon hybridization influences anharmonicity near magic angle, we conducted temperature-dependent Raman experiments on $\sim$1$^{\circ}$ t\mbox{-}BLG and Bernal-stacked bilayer graphene samples over a temperature range of 6 K to 300 K. Figure \ref{Fig6} compares the spectra of $\sim$1$^{\circ}$ t\mbox{-}BLG and Bernal-stacked bilayer graphene at both low and room temperatures. The G$^+$ as well as G$^-$ modes in $\sim$1$^{\circ}$ t\mbox{-}BLG exhibit a more pronounced red-shift compared to Bernal-stacked bilayer graphene as the temperature increases from low to room temperature.

    The temperature-dependent Raman shift of the G mode is primarily attributed to cubic anharmonic interactions involving three-phonon processes. The shift in phonon frequency with temperature arises mainly from the self-energy variation due to direct phonon-phonon coupling and the effect of thermal expansion \cite{bonini2007phonon, lin2011anharmonic, tristant2019phonon}. In this study, the contribution of thermal expansion to the phonon frequency shift is negligible because the temperature range is limited from 6 K to room temperature (300 K). Anharmonic interactions allow the optical phonon to decay into two phonons, a process known as cubic anharmonic decay, where a phonon decays into two phonons, conserving total energy and momentum \cite{bonini2007phonon, lin2011anharmonic, zhu2021temperature}. To quantify the impact of anharmonicity on the temperature-dependent phonon frequency in graphene samples, we use the simplest Klemens model for three-phonon scattering, where an optical phonon of frequency $\omega_{\circ}$ decays into two phonons of equal frequency $\omega_{\circ}/2$:
    \begin{equation}
        \omega(T) = \omega_{\circ} - A \left[ 1 + \frac{2}{e^{\hbar\omega_{\circ}/2KT}-1} \right]
        \label{anharmonic-eq}
    \end{equation}
    Here, $K$ is the Boltzmann constant, $T$ is the temperature, and $A$ is the anharmonic coefficient (with positive sign) \cite{lin2011anharmonic, klemens1966anharmonic}.

    Figure \ref{Fig7} shows the temperature-dependent frequency shifts of the G$^+$ and G$^-$ modes in $\sim$1$^{\circ}$ t\mbox{-}BLG, as well as the G mode in Bernal-stacked bilayer graphene sample. The anharmonic coefficient A, extracted using Eq. \ref{anharmonic-eq}, of the split G modes in the $\sim$1$^{\circ}$ t\mbox{-}BLG sample (279.9 cm$^{-1}$ and 226.3 cm$^{-1}$) is almost 10 times larger than that in the Bernal-stacked bilayer graphene (24.8 cm$^{-1}$). This remarkable enhancement in phonon-phonon coupling shows that phonon-hybridization plays a predominant role in determining the properties of phonons in the twisted bilayer. This can have profound effect on the thermal conductivity of twisted bilayer graphene.

\section{Conclusion}

    In this work, we explored the Raman spectra of twisted bilayer graphene (t\mbox{-}BLG) across a range of twist angles, particularly focusing on the behavior of the G and 2D modes near the magic angle. Our analysis reveals the splitting of the G mode at twist angles close to 1$^{\circ}$, likely caused by phonon hybridization induced by the moir\'{e} potential. Additionally, we observed broadening of the 2D mode near the magic angle, which correlates with enhanced electron-phonon coupling arising from the flat electronic band formation. Temperature-dependent Raman measurements bring out strong anharmonicity in t\mbox{-}BLG, with much larger anharmonic coefficients in the split G modes of the t\mbox{-}BLG compared to the Bernal-stacked bilayer graphene. These results highlight the impact of twist angle and moir\'{e} potential on the phononic properties of t\mbox{-}BLG and suggest further studies on their potential to modify thermal transport.

\section{Acknowledgements}

    The authors thank Shinjan Mandal and Manish Jain for useful discussions. D.S. acknowledges support from the Ministry of Human Resource Development (MHRD), Government of India, in the form of a scholarship under the Integrated Ph.D. program during the course of this research. A.K.S. thanks the Department of Science and Technology (DST), Government of India for financial support under the National Science Chair Professorship and the Nanomission.

\bibliography{bibliography}

\clearpage
\onecolumngrid

\setcounter{figure}{0}

\appendix
\input{si.tex}

\end{document}

%% file: si.tex
\begin{center}
    \large\textbf{Supporting Information}
\end{center}
\vspace{1cm}

\renewcommand{\thefigure}{S\arabic{figure}}

\section{Raman Spectra of Monolayer Graphene}

    \begin{figure}[h]
        \centering
        \includegraphics[width=0.7\textwidth]{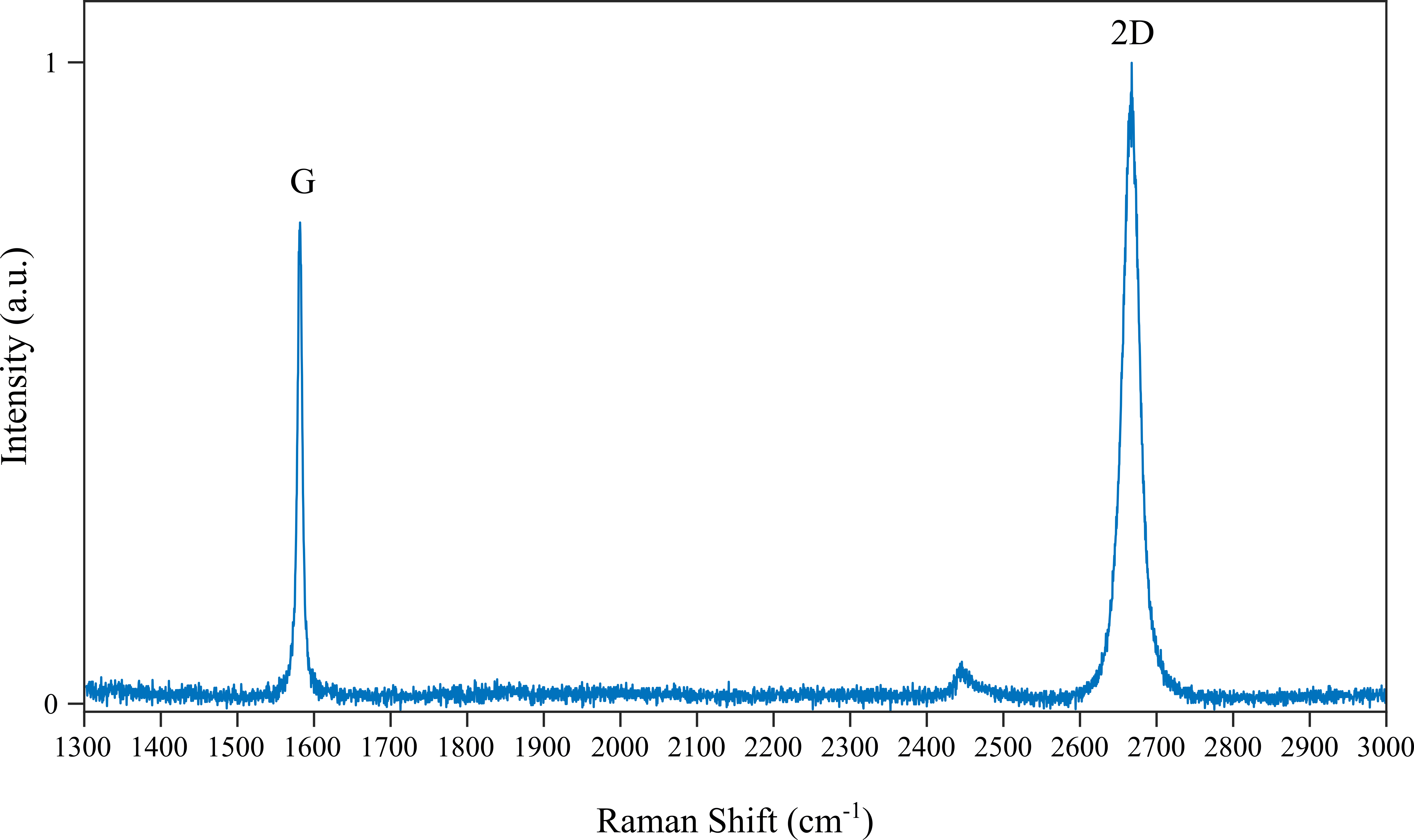}
        \caption{Raman spectrum of monolayer graphene.}
        \label{Gr-Mo}
    \end{figure}

    \noindent Figure \ref{Gr-Mo} shows the Raman spectrum of mechanically exfoliated single-layer graphene (SLG) using the Scotch tape method. The G and 2D modes of graphene appear at approximately 1583 cm$^{-1}$ and 2700 cm$^{-1}$, respectively. The spectrum shows the absence of the D mode ($\sim$1350 cm$^{-1}$), which is a defect-induced mode that appears when translational symmetry is broken. The absence of this peak indicates the high quality of the exfoliated graphene layers.

\newpage
\section{Tear-and-Stack Method}

    \begin{figure}[h]
        \centering
        \includegraphics[width=0.8\textwidth]{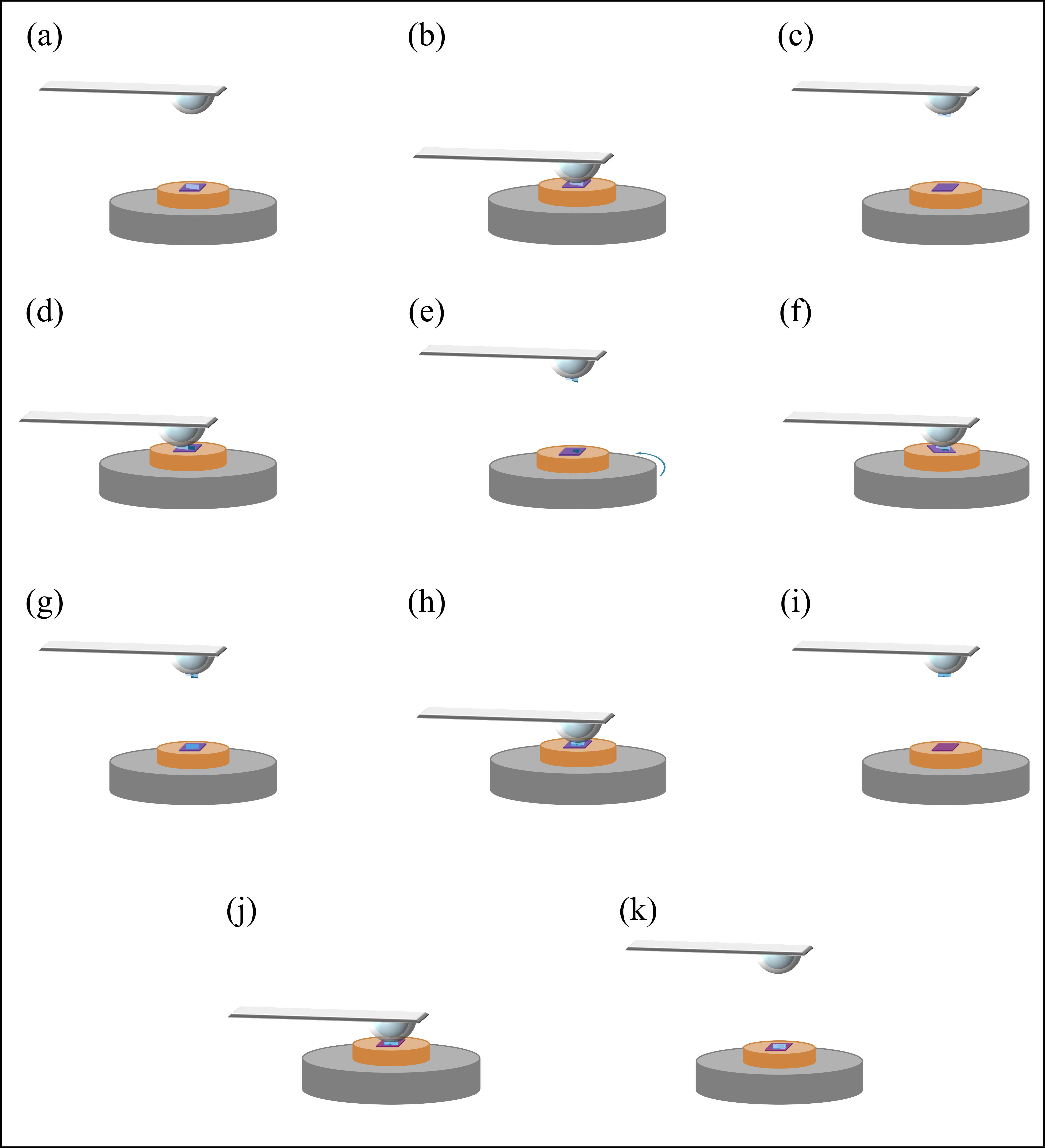}
        \caption{Figures (a)-(k) depict the various stages of the heterostructure fabrication process.}
        \label{fabrication}
    \end{figure}

    The tear-and-stack method for fabricating twisted bilayer samples involves several steps:

    \begin{itemize}
        \item \textbf{Step 1: Preparation and Initial Layer Transfer}
        \begin{enumerate}
            \item[(a)] Start by placing the substrate on a rotating stage and align it with a poly-dimethylsiloxane (PDMS) stamp mounted on a glass slide. As depicted in Figure \ref{fabrication}(a), lower the glass slide towards the exfoliated hexagonal boron nitride (hBN) layer while heating the stage to approximately 80$^\circ$C.
            \item[(b)] Make contact between the stamp and the hBN layer, then reduce the temperature to cool the stage.
            \item[(c)] Carefully lift the glass slide, allowing the thin hBN crystal to adhere to the PDMS stamp.
        \end{enumerate}
    
        \item \textbf{Step 2: Monolayer Alignment and Tearing}
        \begin{enumerate}
            \item[(d)] Place a monolayer of the desired material (e.g., graphene or WSe$_2$) under the stamp for alignment with the hBN. Lower the glass slide again, maintaining the stage temperature at about 60$^\circ$C. Align the hBN to touch half of the monolayer.
            \item[(e)] After a few minutes, pull back the glass slide, tearing the monolayer into two parts: one part remains on the substrate, while the other adheres to the hBN. Rotate the remaining monolayer on the substrate to the desired twist angle.
            \item[(f)] Carefully stack the two parts to form a bilayer with the specified twist angle, ensuring a clean interface for optimal electronic and optical properties.
        \end{enumerate}
    
        \item \textbf{Step 3: Encapsulation}
        \begin{enumerate}
            \item[(g)-(i)] Pick up another hBN layer several atomic layers thick, in a similar manner as in previous steps, to encapsulate the twisted heterostructure and provide protection and stability.
        \end{enumerate}
    
        \item \textbf{Step 4: Final Transfer to Substrate}
        \begin{enumerate}
            \item[(j)-(k)] With the complete heterostructure assembled on the PDMS stamp, introduce a clean silicon (Si) substrate. Transfer the entire heterostructure onto the new substrate by lowering the stamp and heating the stage to around 80$^\circ$C to release the layers.
        \end{enumerate}
    
        \item \textbf{Final Cleaning}
        
        After completing all these steps, the sample is cleaned in chloroform to remove any residual contaminants. This ensures a clean, residual free sample for further experiments. Throughout the process we have taken care so that mechanical stress and contamination are avoided, which can affect the stacking quality, interface cleanliness, and overall sample quality. Clean and precise handling of the 2D material samples is very important to ensure high-quality, defect-free samples for optical and spectroscopic measurements.
    \end{itemize}

\section{FWHM vs Temperature for the Split Peaks of $\sim$1$^{\circ}$ t-BLG}

    \begin{figure}[h]
        \centering
        \includegraphics[width=0.6\textwidth]{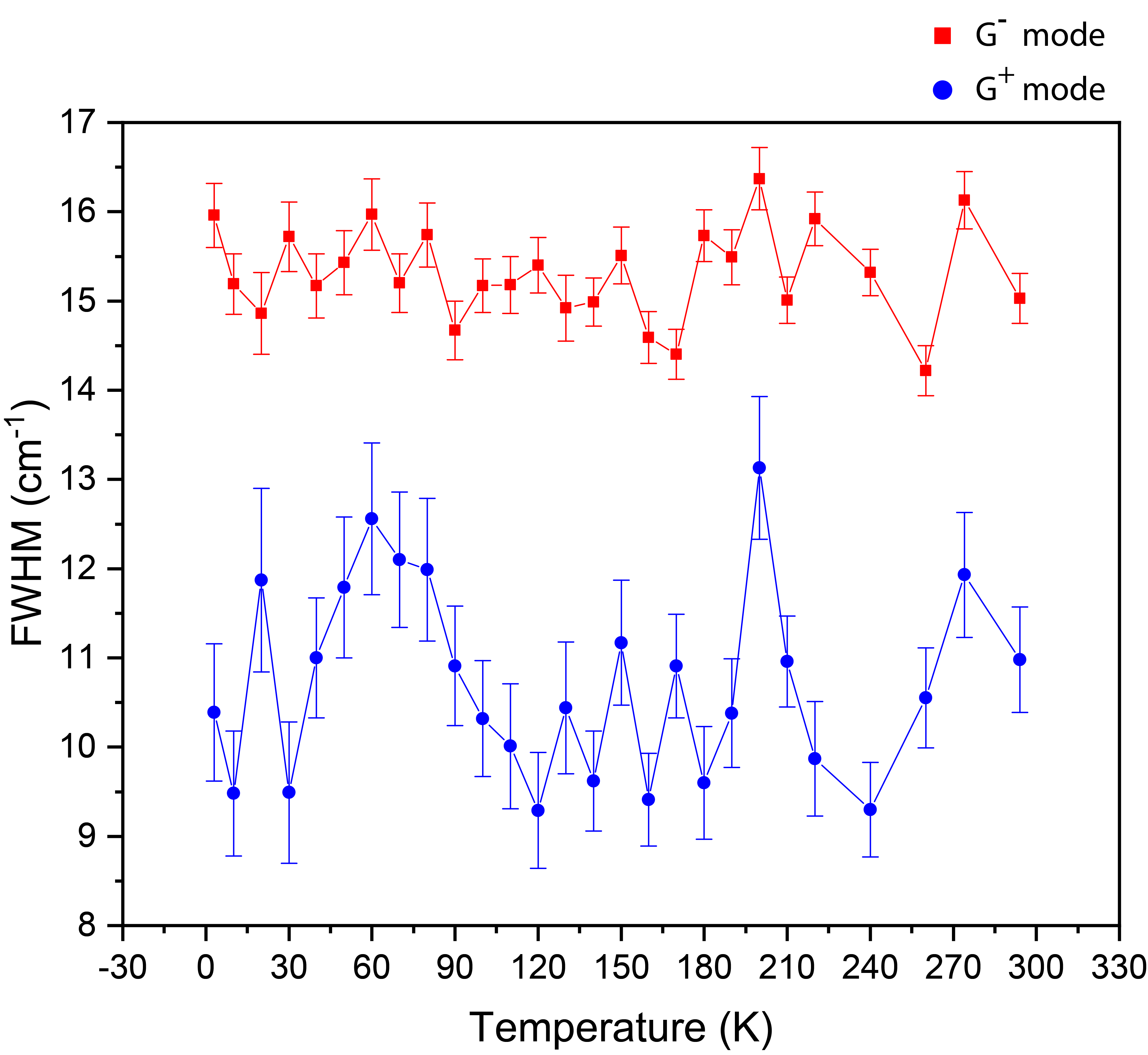}
        \caption{This figure illustrates the variation in the full width at half maximum (FWHM) as a function of temperature for both the G$^-$ (red) and G$^+$ (blue) modes.}
        \label{fwhm}
    \end{figure}

\hspace{1cm}
\section{Relative Intensities vs Temperature for the Split Peaks of $\sim$1$^{\circ}$ t-BLG}

    \begin{figure}[h]
        \centering
        \includegraphics[width=0.6\textwidth]{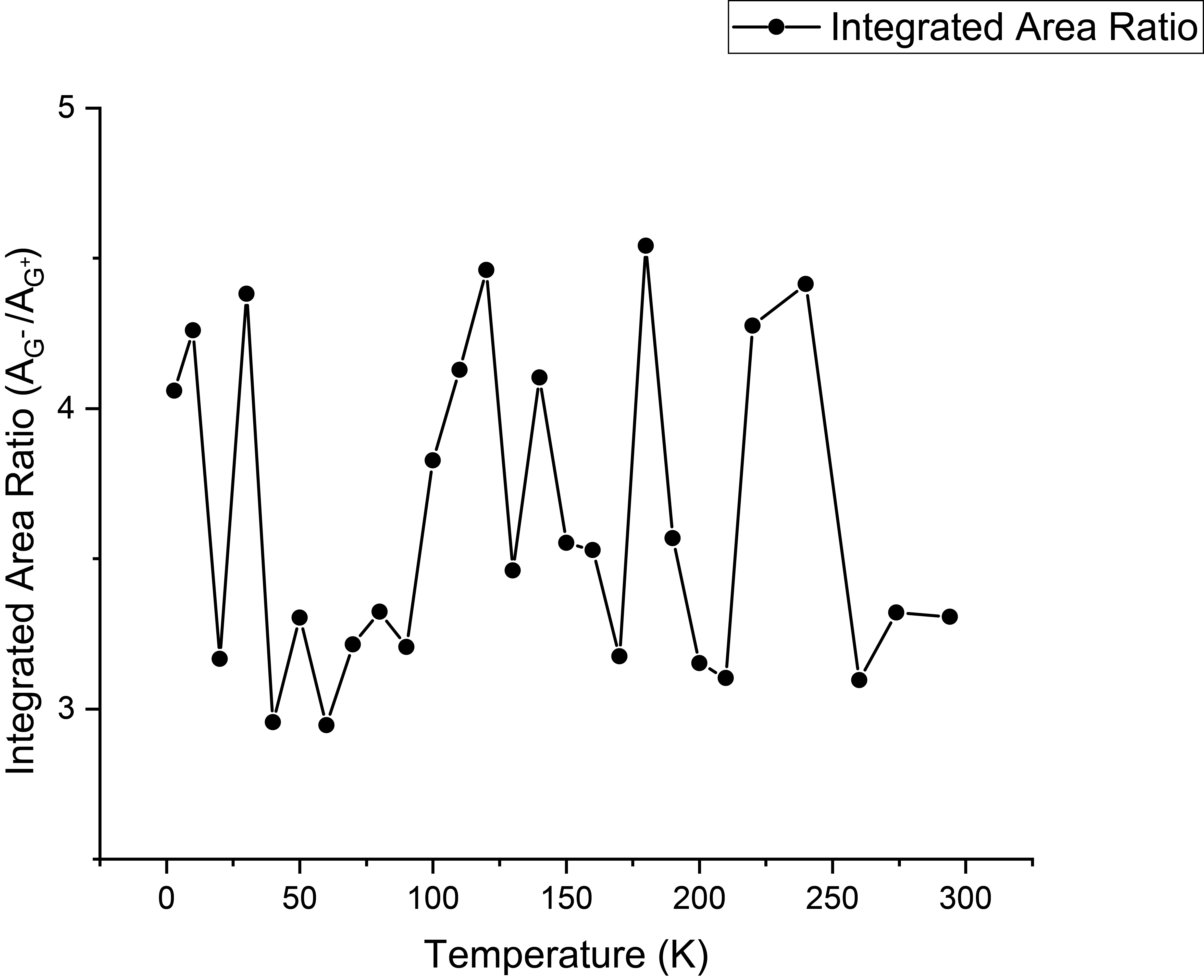}
        \caption{This figure depicts the relative intensity of the G$^-$ and G$^+$ modes, defined as the ratio of integrated intensities (A$_{G^-}$/A$_{G^+}$), as a function of temperature.}
        \label{intensity}
    \end{figure}

\section{Raman Spectra of Monolayer Graphene in the Vicinity of Twisted Region of $\sim$1$^{\circ}$ t-BLG}

    \begin{figure}[h]
        \centering
        \includegraphics[width=0.53\textwidth]{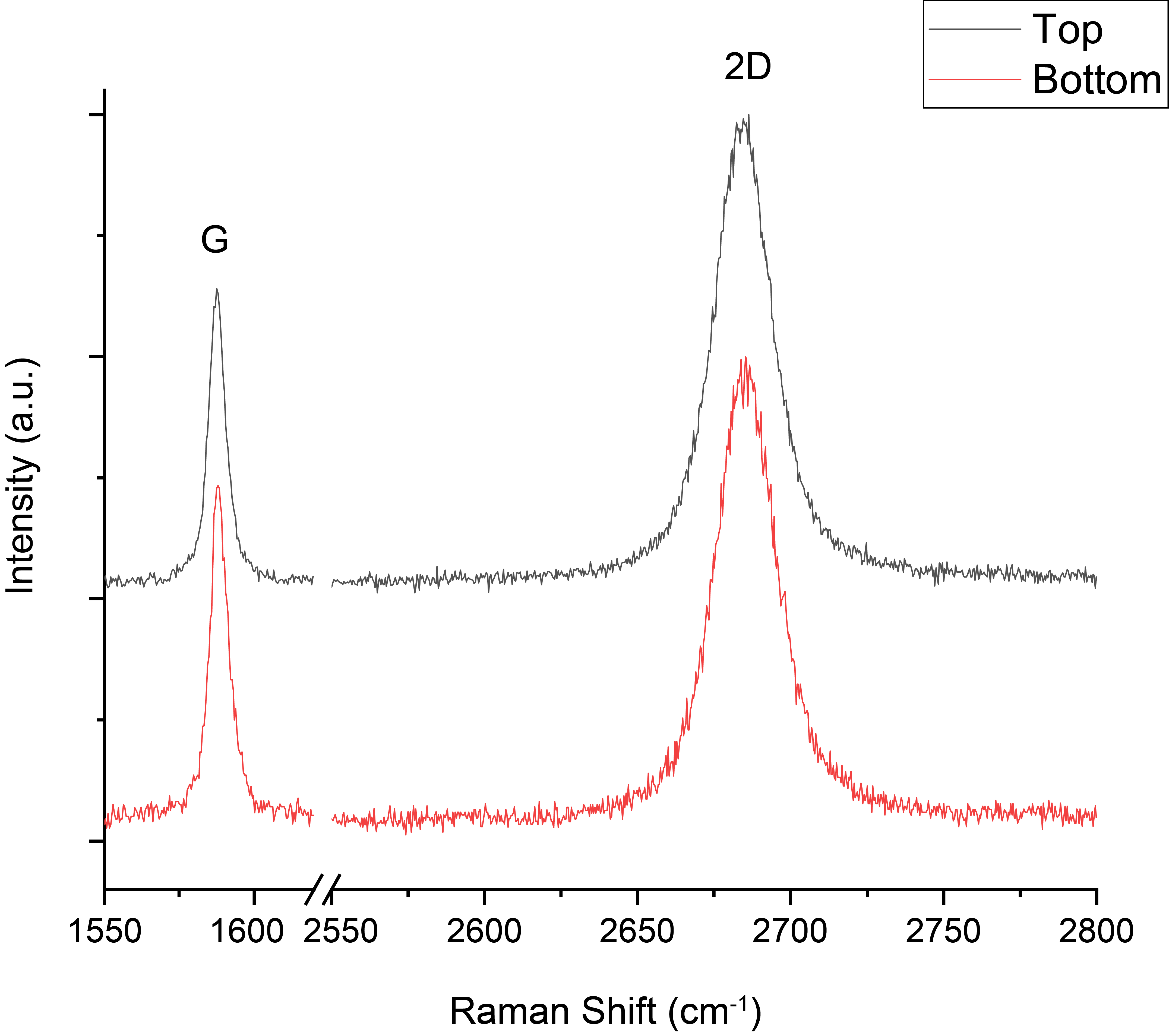}
        \caption{Raman spectra showing the G and 2D modes of both top and bottom monolayer graphene near the twisted region of magic-angle ($\theta_t$ $\sim$ 1$^{\circ}$) twisted bilayer graphene.}
        \label{tb_near-magic}
    \end{figure}